\documentstyle[aps,prb,amsfonts,amssymb,floats,epsf,psfig,pstricks]{revtex}
\begin{document}
\draft
\wideabs{
\title{Doping-dependent study of the periodic Anderson model
in three dimensions}

\author{Thereza Paiva,\protect\cite{address_TP}}
\address{Instituto de F\'\i sica, Universidade Federal do Rio de Janeiro,
Cx.P. 68.528, 21945-970 Rio de Janeiro, RJ, Brazil}
\author{G\"{o}khan Esirgen,\protect\cite{address_GE} and Richard~T. Scalettar}
\address{Department of Physics, University of California, Davis,
California 95616-8677}
\author{Carey Huscroft}
\address{Hewlett-Packard Company, Roseville, California 95747-5200}
\author{A.~K. McMahan}
\address{Lawrence Livermore National Laboratory, University of California,
Livermore, California 94550-9234}
\date{\today}
\maketitle
\begin{abstract}

We study a simple model for $f$-electron systems,
the three-dimensional periodic Anderson model,
in which localized
$f$ states hybridize with neighboring $d$ states.  The $f$ states have
a strong on-site repulsion which suppresses
the double occupancy and can lead to the formation
of a Mott-Hubbard insulator.  When the
hybridization between the $f$ and
$d$ states increases, the effects of these
strong electron correlations gradually
diminish, giving rise to interesting phenomena on the way.
We use the exact quantum Monte Carlo, approximate diagrammatic
fluctuation-exchange approximation, and mean-field Hartee-Fock methods
to calculate
the local moment, entropy, antiferromagnetic structure factor, singlet
correlator, and internal energy as a
function of the $f$-$d$ hybridization for various dopings.
Finally, we discuss the relevance of this work to the volume-collapse
phenomenon experimentally observed in $f$-electron systems.
\end{abstract}
}

\section{INTRODUCTION}
\label{sec_int}

The periodic Anderson model (PAM) describes the qualitative
physics of solids in which a set of localized ($f$)
orbitals having a strong on-site repulsion
hybridizes with another set of
noninteracting conduction ($d$) orbitals.\cite{GENPAM}
The basic magnetic features encapsulated in the PAM include
the formation of moments on the localized orbitals due to
the suppression of double occupancy by the repulsion.
The magnetic nature of the ground state is determined by the strength
of the hybridization between the localized and conduction
electrons. \cite{doniach} If the hybridization is small, then the
magnetic ordering of the localized moments takes place through the
establishment of a modulation of the conduction-electron spin density,
an indirect Ruderman-Kittel-Kasuya-Yosida (RKKY) coupling. \cite{RKKY}
For strong hybridization the screening of magnetic the moments of
the correlated electrons by conduction electrons (Kondo screening) is 
expected.
While some of these general ideas are well accepted,
others are still under study. \cite{gubernatis2003}  For example, 
Noziere's
notion of 'exhaustion',\cite{NOZIERES}
that only the conduction electrons
near the Fermi surface can participate in singlet formation and
hence there are not enough of them to form singlets with all the
local spins, has received considerable recent 
scrutiny.\cite{RECENTEXHAUST}
Related to this is the effectiveness of the compensating
cloud around each localized orbital.
The size of this cloud has been a subject of controversy. Arguments based 
on scaling and renormalization group ideas for the single-impurity Kondo 
point to a cloud with length $\xi_K \sim v_F/T_K$, where $v_F$ is
the Fermi velocity and $T_K$ is the Kondo temperature 
\cite{Affleck98,Affleck2001}. Since Kondo temperatures are normally of the 
order of tens of degrees, this leads to a large Kondo cloud of the order 
of thousands of lattice spacings $a$. On the other hand,  similar 
arguments for the single-impurity multichannel Kondo model 
have led Gan \cite{gan} to find $\xi_K \sim a$.
This cloud has never been observed experimentally and Knight shift
experiments from Boyce and Slichter \cite{boyce} were interpreted to
indicate the absence of a large cloud.

The charge excitations of the PAM also show interesting features.
As in the one-band Hubbard model, a
strong repulsion on the localized orbitals makes them
exhibit the characteristics of a Mott insulator,
including a suppression of the $f$ density of states
and a vanishing $f$-electron compressibility.\cite{ourprl2}
The density of states can also exhibit a Kondo resonance as it evolves 
to a single broad peak at weak coupling.\cite{ourprl2}


While many of these general ideas are well accepted,
much of the quantitative physics of the PAM
remains open.  Only relatively recently has the phase diagram
of the PAM been studied by non-mean-field
approaches like the determinant
quantum Monte Carlo\cite{qmcpam,cpqmcpam} (QMC) and dynamical
mean-field theory\cite{DMFTPAM} (DMFT)\@.
The former provides an exact treatment
of correlations on finite lattices but
is computationally demanding and therefore has so far
been restricted in its applications to the particle-hole symmetric point,
where both the $f$ and $d$ orbitals are half filled.
The DMFT approach works explicitly in the thermodynamic limit
and hence can observe finite-temperature
phase transitions without recourse to finite-size-scaling techniques,
but it involves the approximation of ignoring the momentum dependence
of the self-energy.

There are several reasons to believe
that the removal of the restriction of half filling
might reveal different physics.
First, half filling is the most optimal electron
density for antiferromagnetism.  Since a number of
the potential systems for which the PAM has been suggested as an
appropriate Hamiltonian do not exhibit such magnetic correlations,
it is desirable to study the properties of the doped system,
for which antiferromagnetism is
considerably suppressed.

Second, the noninteracting bands have
special features which are unique to half filling.
For example, in the absence of interactions, the canonical choice
of a momentum-independent hybridization
$V_{\bf k}=V$, which corresponds to a hybridization
of the localized orbitals with the conduction orbitals
on the same site, gives rise to a band insulator at half filling.
Meanwhile, the choice
$V_{\bf k}=-2V(\cos k_{x}+\cos k_{y}+\cos k_{z})$, which, instead,
hybridizes the localized
orbitals with the conduction orbitals on the nearest-neighbor
sites,\cite{WHY} leads to a metallic state at half filling; yet,
this hybridization $V_{\bf k}$ vanishes
at the half-filled Fermi surface and, therefore, is
also somewhat artificial.
While the addition of strong correlations is expected to
modify the physics greatly, it is still possible that
the properties of the PAM reflect these
special features of the noninteracting limit at half filling.

Finally, the particle-hole symmetric point is a very special one
for analytic theories. For example second-order perturbation theory (SOPT)
in the on-site interaction becomes exact at the {\em strong-coupling\/}
limit, in addition to being exact at the weak-coupling limit.
This odd result is only true at half filling and, indeed, is
only true for models with a single impurity band hybridizing with
a single conduction band.  Studying the PAM off half filling
will therefore give valuable insight into the nature of
analytic approaches to the case of multiband systems,
where this anomalous success of SOPT does not occur
at any filling.

A specific instance where all three issues come into play is the
Anderson impurity model and the PAM as they are applied to the
Cerium (Ce) volume collapse.\cite{ALLEN,johansson,CEVC,andyprb} While
recent QMC studies of the half-filled PAM have lent greater
quantitative understanding of this problem,\cite{ourprl1} the relevance
of this work for the Ce volume-collapse phenomenon would be much better
justified if the same physics could be demonstrated in the doped
system.

In this paper we will present a detailed study of the magnetic,
charge, and thermodynamic properties of the PAM for various band fillings.
We will make use of
the exact QMC, approximate diagrammatic fluctuation-exchange
approximation (FLEX), and mean-field Hartree-Fock (HF) methods.
We will concentrate on the effects of
doping of additional electrons into the system for parameter regimes
where the number of localized (i.e., $f$) electrons still remains
approximately one.  This is a natural consequence of the Mott-Hubbard
gap in the PAM and similarly is also the experimental situation
in at least one application of interest---the Ce volume-collapse transition.

Our key conclusions are the following:
(1)~The more accurate treatment of correlations provided by QMC and
FLEX shift the HF transition to stronger coupling and decrease its
sharpness.  HF for the PAM gives more accurate energies at low
temperature near strong coupling than weak.
(2)~Antiferromagnetic correlations remain robust for
a range of dopings of the conduction band.
By comparison the one-band Hubbard model
loses its long-range antiferromagnetic
order more rapidly upon doping.\cite{HIRSCH,SRW,COMMENT}
(3)~Anomalies in the singlet correlator, entropy, and internal energy
persist even when the f-d hybridization has substantial values
on the Fermi surface.
They also persist to doping levels where antiferromagnetic correlations
have been eliminated.  However, their sharpness does appear correlated
with the presence of magnetism.
(4)~FLEX captures many of the same qualitative features
exhibited by QMC
in the dependence on the interband hybridization
of the local moment,
antiferromagnetic correlations, entropy, and internal energy,
especially at weak coupling, where the HF internal energy is not
very accurate.  At the
same time, as is well known, FLEX has difficulty reproducing the
Mott-Hubbard gap in the strong-coupling limit.\cite{motthubbardgap}

\section{MODEL AND METHODS}
\label{sec_mod}

\subsection{Periodic Anderson model}
\label{sec_pam}

We will study the grand-canonical Hamiltonian
\begin{eqnarray}
\lefteqn{{\cal H} - \mu {\cal N}=}
\nonumber\\
&&\:\:\:\sum_{\langle {\bf i},\:{\bf j}\rangle,\:\sigma}
\{[-t_{dd}d_{{\bf i} \sigma}^{\dagger} d_{{\bf j} \sigma}^{\phantom{\dagger}}
-t_{fd}(d_{{\bf i}\sigma}^{\dagger} f_{{\bf j}\sigma}^{\phantom{\dagger}}
+ f_{{\bf i}\sigma}^{\dagger} d_{{\bf j}\sigma}^{\phantom{\dagger}})]
+{\rm H.c.}\}
\nonumber\\
&&\:\:\:\:\:\mbox{}+\varepsilon_f \sum_{{\bf i},\:\sigma}n_{f{\bf i}\sigma}
-\mu \sum_{{\bf i},\:\sigma}(n_{d {\bf i} \sigma}+n_{f {\bf i} \sigma})
\nonumber\\
&&\:\:\:\:\:\:\:\:\mbox{}+U_{f}\sum_{{\bf i}}
n_{f {\bf i} \uparrow}n_{f {\bf i} \downarrow}.
\label{Hamiltonian}
\end{eqnarray}
$d_{{\bf i} \sigma}^{\dagger}$ and $f_{{\bf i} \sigma}^{\dagger}$
create $d$ (conduction) and $f$ (localized)
electrons with spin $\sigma$ on the lattice site ${\bf i}$.
We consider a three-dimensional periodic simple-cubic lattice, where
$\langle {\bf i},\:{\bf j} \rangle$
is a sum over nearest-neighbor bonds.
Therefore, this term represents the
nearest-neighbor hybridization integrals (i.e., hoppings)
of the $d$-$d$ ($t_{dd}$) and
$d$-$f$ ($t_{fd}$) orbitals. The remaining one-particle terms
are $\varepsilon_f$, which measures the $f$-orbital energy level
with respect to that of the $d$ orbital
(which was chosen to be zero by convention),
and the chemical potential $\mu$.
The charge density operator on site ${\bf i}$ in orbital $d$ 
with spin $\sigma$ is
$n_{d {\bf i} \sigma}=
d_{{\bf i} \sigma}^{\dagger} d_{{\bf i} \sigma}^{\phantom{\dagger}}$
and similarly for $n_{f {\bf i} \sigma}$. Finally, $U_f$
describes an effective on-site repulsion between the
localized $f$ electrons.

When the hopping term is Fourier transformed, the resulting expression
is $\sum_{{\bf k},\:\sigma}
[\varepsilon_{\bf k}
d_{{\bf k} \sigma}^{\dagger} d_{{\bf k} \sigma}^{\phantom{\dagger}}
+V_{\bf k}
(d_{{\bf k}\sigma}^{\dagger} f_{{\bf k}\sigma}^{\phantom{\dagger}}
+ f_{{\bf k}\sigma}^{\dagger} d_{{\bf k}\sigma}^{\phantom{\dagger}})]$,
with $\varepsilon_{\bf k}=-2t_{dd}(\cos k_{x}+\cos k_{y}+\cos k_{z})$
and $V_{\bf k}=-2t_{fd}(\cos k_{x}+\cos k_{y}+\cos k_{z})$.
The particular choice of $\varepsilon_{f}=-U_{f}/2$ and $\mu=0$ results
in $\langle n_{d}\rangle=\langle n_{f}\rangle=1$
($\langle n_{d}\rangle \equiv \sum_{\sigma}\langle
n_{d {\bf i} \sigma}\rangle$ etc.\,\@.).
This is the so-called ``symmetric'' PAM,
in which both orbitals are half
filled regardless of the choice of all the other parameters
and temperature.

We will express all energy units in this paper in terms of
$t_{dd} \equiv 1$. We will also set
$U_{f}/t_{dd}=6$ throughout the paper and vary $t_{fd}/t_{dd}$ for
several fillings and $\varepsilon_{f}/t_{dd}$.  For $\alpha$- and
$\gamma$-Ce respectively, calculated values for the effective- hopping
and Coulomb-repulsion parameters in eV are $t_{dd}\approx 0.90$ and
$0.74$, $t_{fd}\approx 0.19$ and $0.14$, $U_{f}\approx 5.7$ and
$5.9$.\cite{CEVC}  Our choices of the model parameters are
therefore in rough agreement with the values for Cerium.
Finer tuning would not be useful given the approximations inherent in 
using a two-band model.\cite{LDADMFT}  


\subsection{Determinant quantum Monte Carlo}
\label{sec_qmc}

The determinant-QMC method that we employ is standard and
has been described many times in the literature.\cite{DETQMC}
Its key features are an exact treatment of the interactions via
a path-integral representation of the partition function and the
decoupling of the interaction via the introduction of a
Hubbard-Stratonovich field, whose possible
configurations are then summed over stochastically.
The approach is able to study several hundred interacting
electrons, much greater than other exact approaches like
diagonalization, but still representing small lattices, especially
in three dimensions.  This is an especially serious consideration
in determining the presence or absence of long-range order.
Local quantities like the internal energy and near-neighbor spin
correlations have much smaller finite-size effects.
The simulations reported here are on lattices of $64$ ($4\times 4\times 4$)
spatial sites with fillings ranging from
$\langle n \rangle =2.0$ to
$\langle n \rangle =3.6$ electrons per site, where $\langle n\rangle
\equiv \langle n_{d}\rangle+\langle n_{f}\rangle$.

Besides finite spatial-lattice size,
a crucial bottleneck in the determinant QMC is the ``sign problem,''\cite{LOH}
which, roughly speaking for this application,
restricts the temperatures accessible to simulation
to those greater than $T \sim W/50$, where $W$ is the noninteracting
bandwidth.

\subsection{Fluctuation-exchange approximation}
\label{sec_flex}

We will also study the PAM using the
fluctuation-exchange (FLEX) approximation.\cite{FLEX}
The FLEX approximation can be motivated in various different ways.
It can be thought of as an extension of the Hartree-Fock theory,
in which the bare interaction is self-consistently
screened by the exchange of
electron-hole and electron-electron (two-particle) fluctuations.
Or it can be viewed as a conserving approximation within the
framework of Baym-Kadanoff's
generating-functional formalism.\cite{baymkadanoff}
Another view is as an extension of the random-phase approximation (RPA), in
which the two-particle
propagators of RPA are self-consistently included in the self-energy.
Yet another, but a very elegant one, is that FLEX is the first step
in the parquet-approximation's\cite{PARQUET}
two-particle-self-consistency scheme.

FLEX suffers from lack of the two-particle self-consistency found
in the parquet approximation,
which renormalizes the bare vertices (which are used to
calculate the fluctuation propagators) with vertex corrections.
Although it is possible to formulate FLEX to work
in spontaneously-broken-symmetry
states (with an anomalous self-energy), such as a magnet or
superconductor, FLEX as commonly practiced rarely does so.
The only such FLEX calculations to this date include those for a
superconductor.\cite{SCFLEX}
For previous results using FLEX for Hubbard-like
models, along with comparisons with QMC, see Ref.~\onlinecite{MOREFLEX}.

The normal-state FLEX solutions of Hubbard-like models are also known to
give the incorrect strong-coupling limit. It is not known
if the broken-magnetic-symmetry
solutions may give the right limit for main thermodynamic
quantities at strong coupling.

The FLEX approximation used in this paper incorporates all four channels
(density, magnetic, singlet, and triplet fluctuations). In 
order to satisfy certain symmetries, all these four channels are needed.
In fact we also tested FLEX without using the particle-particle
fluctuations, and it turned out that the resulting
accuracy is lower. 
We also use
the numerical renormalization-group method on the Matsubara
frequencies to accelerate evaluation of the
FLEX self-energies and fluctuation propagators.\cite{paorg}

\section{RESULTS AND DISCUSSION}
\label{sec_res}

In this section we will show how the properties of the PAM evolve
from strong to weak coupling at several different fillings.
We begin by describing the orbital occupations, showing how the charge
is transferred with the temperature.  This helps establish relevant temperature
scales in these simulations.  We then show short-range magnetic
correlations---measures of the local moment and singlet formation---before 
turning to the issue of long-range antiferromagnetism.
The section concludes with results for the entropy and internal energy
and a general discussion of the relation of their behavior
with magnetism.

As mentioned in Section~\ref{sec_pam},
the PAM Hamiltonian in Eq.~(\ref{Hamiltonian}) with $\varepsilon_{f}=-U_{f}/2$
and $\mu=0$ has particle-hole symmetry---$\langle n_{d}\rangle=
\langle n_{f}\rangle=1$---i.e., both the conduction and localized
orbitals are half filled, regardless of all other parameters.
We will compare the properties of the PAM at two electron
densities with the behavior at half filling.
The density $\langle n \rangle=2.2$
is only lightly doped away from half filling and allows us to examine
the rapidity of the destruction of the antiferromagnetism.
The density $\langle n \rangle=2.6$, as we shall see,
is large enough that all traces of long-range order have been suppressed.
In both these cases, we have adjusted the $f$ site energy to maintain
$\langle n_f\rangle\sim 1$.

Finally,
we also present a few results for
an extreme-filling case, in which $\langle n_d\rangle\sim 1.8$ and
$\langle n_f\rangle\sim 1.8$,\@ where the effect of
electron-electron correlations are shown to be minimal, as is expected
since this is equivalent
to a very low electron density by a particle-hole transformation.

\subsection{Orbital occupations}
\label{sec_orbocc}

\begin{figure}[hbtp]
\psfig{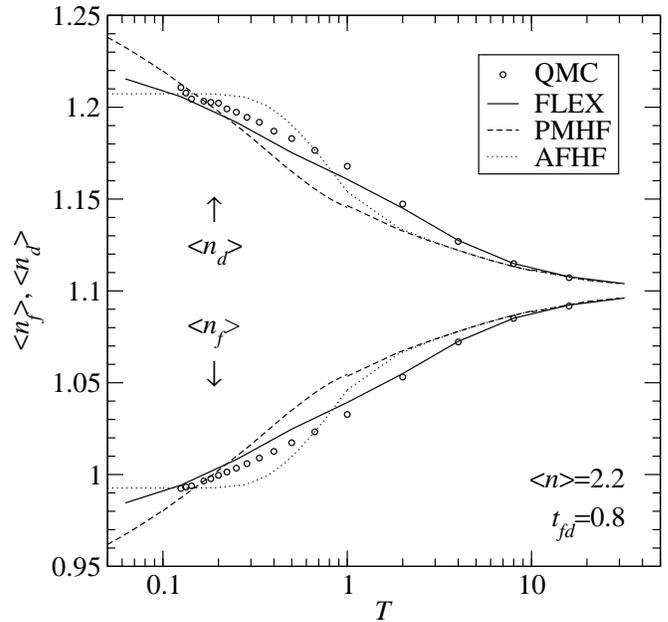}
\vskip 8 pt
\caption{The dependence of the $f$ and $d$ orbital
occupations on the temperature $T$ at $t_{fd}=0.8$ for
$\langle n \rangle =2.2$.
Other parameters are $t_{dd}=1$ and $U_{f}=6$ (throughout this paper),
and $\varepsilon_{f}=-2.785$. Also, the lattice sizes in
this paper are
all $4\times 4\times 4$ unless otherwise specified.
Results for QMC, FLEX, paramagnetic
Hartree-Fock (PMHF), and antiferromagnetic Hartree-Fock (AFHF) are
shown.}
\label{fig_nfndT}
\end{figure}

We begin by examining the evolution of
the local orbital density with
the temperature $T$ in Fig.~\ref{fig_nfndT}.
At high $T$ the orbitals are equally filled
since energies like $\varepsilon_{f}$, which distinguish the
orbitals from each other, are washed out by large thermal fluctuations.
As $T$ is lowered, the occupations shift.
By $T \sim 0.125$ ($\beta\equiv 1/T\sim 8$),
much, but not all, of the charge transfer from $f$ to $d$ orbitals
is complete for this value of $t_{fd}=0.8$.
For smaller $t_{fd}$ the occupations reach their asymptotic
ground state 
values at higher temperatures owing to the presence of
a larger Mott gap.
Figure \ref{fig_nfndT}
also shows the data for the Hartree-Fock solutions to the problem,
demonstrating that these techniques are in qualitative agreement with the
exact QMC calculations for the temperature dependence of the orbital
densities, despite the fact that Hartree-Fock predicts magnetic order below
$T \sim U_{f}/4$ ($=1.5$ for $U_{f}=6$),
well above any possible QMC transition temperature.
The FLEX results are in even better qualitative agreement with the
exact QMC results in the entire temperature range shown.

\subsection{Local moment}
\label{sec_localmoment}

\begin{figure}[hbtp]
\psfig{figure=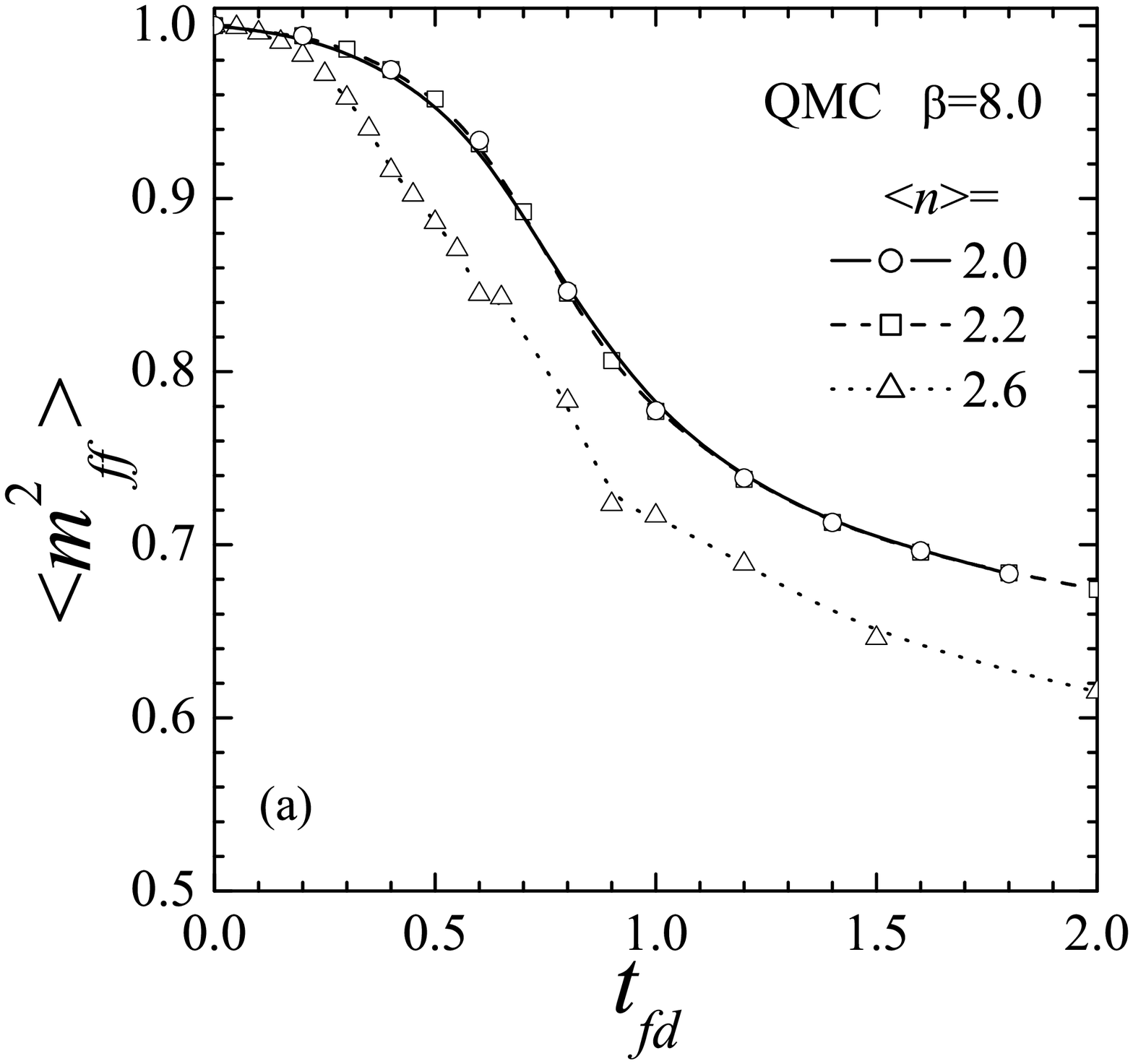,width=330 pt,angle=0}
\vskip 8 pt
\psfig{figure=fig2b.eps,width=246 pt,angle=0}
\vskip 8 pt
\caption{(a)~The dependence of the QMC local moment on
the hybridization $t_{fd}$ at low temperature $T=1/8$ 
($\beta=1/T=8$) for
$\langle n \rangle=2.0$, $2.2$, and $2.6$. $\varepsilon_{f}=-3.0$,
$-2.785$, and $-2.6$, respectively. Other parameters are the same
as in Fig.~\protect\ref{fig_nfndT}. (b)~The FLEX results for the same
parameters for three-different lattice sizes, showing 
finite-size effects are small.
The quantity shown is the $\Delta {\bf R}={\bf 0}$, $\Delta \tau
=0$ component of the $f$-orbital magnetic susceptibility, which is the
same as the QMC local moment shown in (a).\cite{vertex}
The QMC and FLEX results are very similar.
}
\label{fig_localmoment}
\end{figure}

Figure \ref{fig_localmoment}(a) shows
the dependence of the $f$-orbital moment,
$\langle m_{ff}^2 \rangle\equiv
\langle (n_{f{\bf i}\uparrow}- n_{f{\bf i}\downarrow})^2 \rangle =
\langle n_{f} \rangle - 2
\langle n_{f{\bf i}\uparrow} n_{f{\bf i}\downarrow} \rangle$,
on the hybridization $t_{fd}$ at low temperature.
Here data for several dopings are shown.  For all three cases,
$\langle m_{ff}^2 \rangle$
is fully formed at weak hybridization, and,
as expected, it is reduced as $t_{fd}$ increases and
the system becomes less strongly correlated. Even for relatively large
$t_{fd}$, however,
$\langle m_{ff}^2 \rangle$
is significantly enhanced over its uncorrelated value of
$\langle m_{ff}^2 \rangle=1/2$.
For example, the value
$\langle m_{ff}^2 \rangle= 0.8$
at $t_{fd}=1$ and $\langle n \rangle=2.2$
corresponds to a double occupation of
$\langle n_{f{\bf i}\uparrow} n_{f{\bf i}\downarrow} \rangle =0.10$, a value
reduced by a factor of two-and-a-half
from the result in the absence of correlations,
$\langle n_{f{\bf i}\uparrow} n_{f{\bf i}\downarrow} \rangle=
\langle n_{f{\bf i}\uparrow}\rangle
\langle n_{f{\bf i}\downarrow} \rangle=1/4$.
Although the moment is
reduced as $t_{fd}$ increases, even at large $t_{fd}$, the moment
is well enough formed to allow for interesting magnetic ordering to occur.
For $U/t=4$, the half-filled one-band Hubbard model has
$\langle m^2 \rangle=0.77$ at low $T$, a value comparable to the smaller of
those in Fig.~\ref{fig_localmoment}(a);
yet, the one-band Hubbard model still exhibits clear
long-range antiferromagnetic order for this $U/t$.
Later in this paper, we will examine antiferromagnetism in the PAM.

In Fig.~\ref{fig_localmoment}(b) the FLEX results for
the local and instantaneous
component of the magnetic susceptibility,
$\chi_{m}(\Delta {\bf R}={\bf 0},\:\Delta\tau=0)$,
are shown for different lattice sizes.\cite{vertex}
This quantity is the equivalent of the local moment.
As depicted in the
figure, the FLEX and QMC results are very similar, although FLEX somewhat
underestimates the local moment.  The finite-size effects
are very small, as would be expected for a local quantity.

\subsection{Singlet correlator}
\label{sec_singletcorr}

We will now turn to the issue of the Kondo screening of the local moments.
Here we probe the
formation of local Kondo singlets, in which one conduction electron and 
one localized electron in nearest-neighbor sites form a bound singlet 
state. In order to do that we 
use the singlet correlator, which is the nearest-neighbor interband 
spin-spin correlation function and is  defined by
$c_{fd}={1 \over 3} \sum_{\bf j}^{nn}\langle \vec S_{f{\bf i}} \cdot \vec
S_{d{\bf j}} \rangle$,
where $\vec S_{f{\bf i}} =  \left(\matrix{f^{\dagger}_{{\bf i}\uparrow} &
f^{\dagger}_{{\bf i}\downarrow}}\right)
\vec\sigma\left(\matrix{f_{{\bf i}\uparrow}\cr f_{{\bf 
i}\downarrow}}\right)$ 
and likewise for $\vec S_{d{\bf j}}$.
This definition of $c_{fd}$ is a natural extension of the dot product of
the local electron spin and conduction electron spin on the same lattice 
site which is widely used to measure singlet formation within a PAM with 
on-site $fd$ hybridization.  As discussed above, our model has an 
intersite $fd$ hybridization which leads to singlet formation of the local 
electron spin with conduction electrons on neighboring sites. 
\cite{ourprl1}

\begin{figure}[hbtp]
\psfig{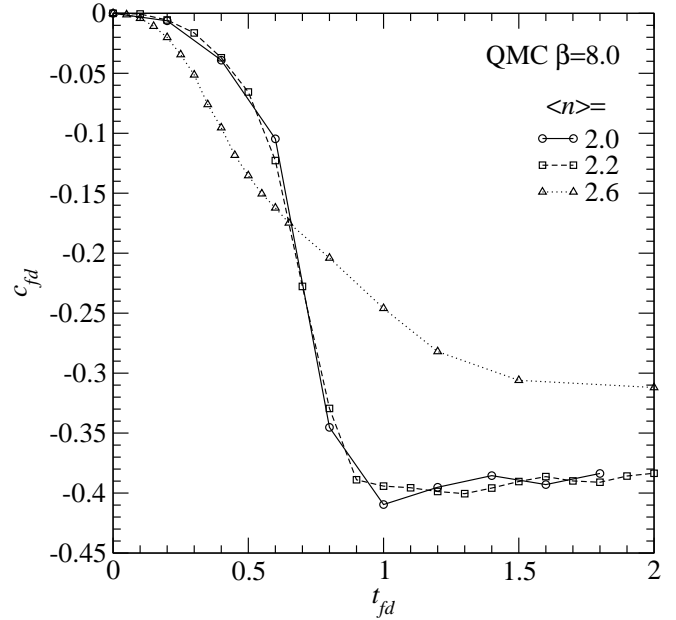}
\vskip 8 pt
\caption{The hybridization dependence of the singlet correlator, which
measures the degree of the local antiferromagnetism
between an $f$ moment and the conduction electron spins
on the neighboring sites.
Parameters are the same as in Fig.~\protect\ref{fig_localmoment}(a).}
\label{fig_singletcorr}
\end{figure}

In Fig.~\ref{fig_singletcorr} we show the dependence of 
$c_{fd}$ on the hybridization $t_{fd}$ for different fillings,
calculated by QMC.
There is a qualitative difference between the $\langle n\rangle=2.6$ case,
and the $\langle n\rangle=2.0$ and $2.2$ cases. In the latter there is a
sharp onset in the singlet correlator $c_{fd}$ near $t_{fd}=0.6$. This was
suggested to be intimately related to the Kondo volume-collapse scenario
in our previous work for $\langle n\rangle=2.0$.\cite{ourprl1}
This onset moves to smaller $t_{fd}$ and
becomes much less sharp at $\langle n\rangle=2.6$.
The $\langle n\rangle=2.0$ and $2.2$ cases are almost identical,
as will be seen throughout this paper. Thus the fact that our choice of
hybridization vanishes at the $\langle n \rangle=2.0$ Fermi
surface as noted earlier---but not at the $\langle n \rangle=2.2$ Fermi
surface---shows that the former artificiality has no impact on the
conclusions in Ref.~\onlinecite{ourprl1}.

Let us 
call  $T_K$ the temperature where the local singlets are formed. We expect 
$T_K(t_{fd})$ to increase with $t_fd$. \cite{doniach}
The behavior of the singlet correlator in Fig. \ref{fig_singletcorr}
can then be explained:  the temperature used in the simulations ($T=1/8$) 
crosses $T_K$.  As the crossing  point depends on band filling 
we observe different onsets, as discussed above. The crossing of
$T$ and $T_K$ also explains why $c_{fd} \to 0$ 
for small $t_{fd}$.


\subsection{Antiferromagnetism}
\label{sec_af}

\begin{figure}[hbtp]
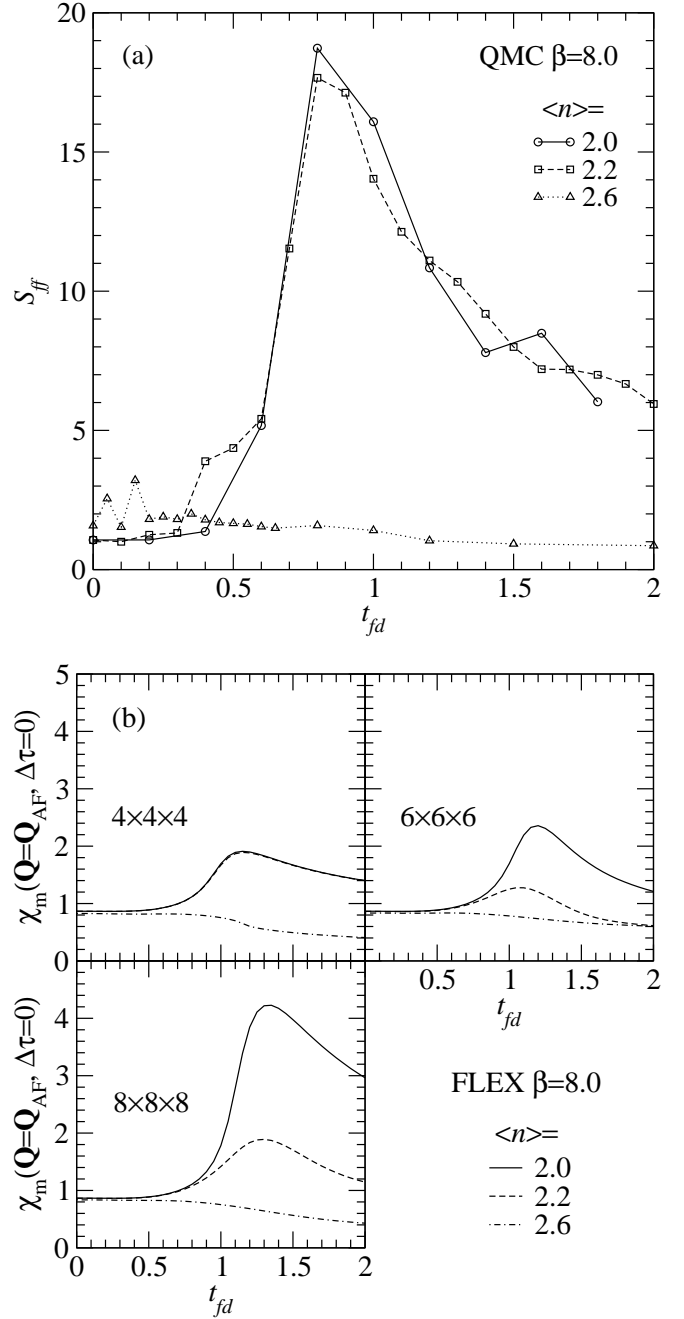

\psfig{figure=fig4a.eps,width=246 pt,angle=0}
\vskip 8 pt
\psfig{figure=fig4b.eps,width=246 pt,angle=0}
\vskip 8 pt
\caption{(a)~The dependence of the 
antiferromagnetic structure factor on the hybridization $t_{fd}$,
as computed with QMC.
Parameters are the same as in Fig.~\protect\ref{fig_localmoment}(a).
(b)~The FLEX results for the same
parameters for three-different lattice sizes.
The quantity shown is the
${\bf Q}={\bf Q}_{\rm AF}=(\pi,\pi,\pi)$, $\Delta \tau=0$
component of the $f$-orbital magnetic susceptibility, which is the
same quantity as in (a).\cite{vertex}
The QMC and FLEX results
are very similar. Note the finite-size scaling
of the FLEX results.
That antiferromagnetism persists at large $t_{fd}$,
when singlet formation is expected in the case of on-site hopping
between $f$ and $d$ orbitals, reflects our
near-neighbor choice of $f$-$d$ hybridization.
Antiferromagnetism is clearly absent for all
$t_{fd}$ at $\langle n \rangle=2.6$.}
\label{fig_sf}
\end{figure}

Another possible phenomenon for the PAM is antiferromagnetic
ordering of the moments. Fig.~\ref{fig_sf}(a) shows
QMC results for the equal time antiferromagnetic
structure factor $S_{ff}$,
the sum of the staggered correlations of the local $f$ spins
across the lattice, 
$S_{ff}(T, {\bf q}=(\pi, \pi, \pi)) = {2\over 3N}  \sum_{{\bf i,j}}
(-1)^{{\bf i}+{\bf j}} \vec S_{f{\bf i}} \cdot \vec S_{f{\bf j}}$.
This structure factor is normalized in such a way that if there
is only short range magnetic order, $S_{ff}$ is of order unity, whereas
if there are long range correlations $S_{ff}$ is of order the
lattice size (and hence diverges in the thermodynamic limit).

In Fig.~\ref{fig_sf}(b) we show FLEX results
for the instantaneous
antiferromagnetic susceptibility $\chi_{m}$.
Just as the instantaneous, local susceptibility
is the FLEX equivalent of the local moment, $\chi_m$ is
the FLEX analog of the QMC structure factor.\cite{vertex}
We will discuss the issue of long-range order below.
However before doing so, we emphasize 
the key observation that such order
for $\langle n \rangle = 2.6$ is clearly completely suppressed.

In interpreting these results further, and in particular the 
possible occurrence of long-range order at
$\langle n \rangle = 2.0$ and $\langle n \rangle = 2.2$, it is
important to emphasize that the Ne\'el temperature depends on
$t_{fd}$.  As in the three-dimensional single-band Hubbard model,
$T_{N}$ is low at small $t_{fd}$, rises to a maximum, and then
falls again at large $t_{fd}$. 
The important point is that a
set of simulations at fixed temperature, like those of
Fig.~\ref{fig_sf}, can potentially cut across the $T_N$ vs.~$t_{fd}$
curve, yielding the impression of the existence of AF order only in
some intermediate range of $t_{fd}$.  With this in mind, we believe
the fall-off of AF order at small $t_{fd}$ for $\langle n \rangle
= 2.0$ and $2.2$ reflects only that $T_{N}$ has gone below the
simulation temperature.  Meanwhile, the fall-off at large $t_{fd}$
may also reflect this reduction of $T_{N}$ or the competing formation of
Kondo singlets.  We note that in the PAM with on-site hybridization,
one expects singlet formation to usurp long-range magnetic
order completely at large $t_{fd}$.\cite{qmcpam}
Our FLEX results suggest the intersite hybridization is 
different in this regard.  Only a very
detailed study including finite-size scaling can sort out these
different possibilities.


To reemphasize, despite all these complications,
Fig.~\ref{fig_sf} still carries an unambiguous message:
the filling
$\langle n \rangle = 2.6$
has a very different behavior in its magnetic properties:
there is no long-range order.
We have already seen that the singlet correlator still
exhibits a growth with $t_{fd}$ at
$\langle n \rangle = 2.6$.
Below we will see that, in the same way,
the $\langle n \rangle = 2.6$
entropy and energy still behave rather similarly to
cases where the densities are close to half filling.

\subsection{Entropy}
\label{sec_entropy}

The low-temperature entropy, $S(T=T_{\rm low})$, offers a means of
probing Kondo-singlet formation as well as antiferromagnetic
order.  Specifically, the entropy per atom of free spin-$1/2$
moments is $\ln 2$, which can be reduced by the formation of local
singlets in cooperation with the surrounding valence electrons or
by the onset of magnetic order.  Since the $f$ and $d$ subsystems
decouple at small $t_{fd}$, the free-moment regime is realized by
the $f$ electrons in this strong-coupling limit for $T_{\rm low}$
above the Ne\'el temperature $T_{\rm N}$, yet well below some
characteristic temperature of moment formation.  $T_{\rm
low}=0.125$ is a reasonable choice since
Fig.~\protect\ref{fig_localmoment}(a) shows fully formed $f$
moments at small $t_{fd}$ for this temperature, while $T_{\rm
N}\rightarrow 0$ as $t_{fd}\rightarrow 0$.  The entropy of the $d$
subsystem may be exactly calculated for $t_{fd}=0$, and at $T_{\rm
low}=0.125$ has already reached its $T=0$ limit:
$S_d(t_{fd}=0,T=0) = 0.43322$, $0.29462$, and $0.31277$ for
$\langle n\rangle=2.0$, $2.2$, and $2.6$, respectively.  These
nonzero $S_d$ values at $T=0$ are finite-size effects of the
$4\times 4\times 4$ lattice arising from the degeneracies of the
$d$ levels at the Fermi surface, which are a nonnegligible fraction
of states in the Brillouin zone for small lattices.  The $f$
electrons are moved away from the Fermi surface by the Mott-Hubbard
gap and are not affected by such finite-size effects.  We calculate
the entropy for the fully interacting system from

\begin{equation}
S(t_{fd},T) = S_\infty - \int_T^\infty dT^\prime
\frac{1}{T^\prime} \frac{\partial E(t_{fd},T^\prime)}{\partial T^\prime}
\label{entropy}
\end{equation}
\noindent
and
\begin{equation}
S_\infty = 4\ln 4 - \langle n\rangle \ln \langle n\rangle
- (4-\langle n\rangle)\ln (4-\langle n\rangle),
\label{Soo}
\end{equation}
using fits to the QMC energies as described
elsewhere.\cite{ourprl1,2dHub}

\begin{figure}[hbtp]
\psfig{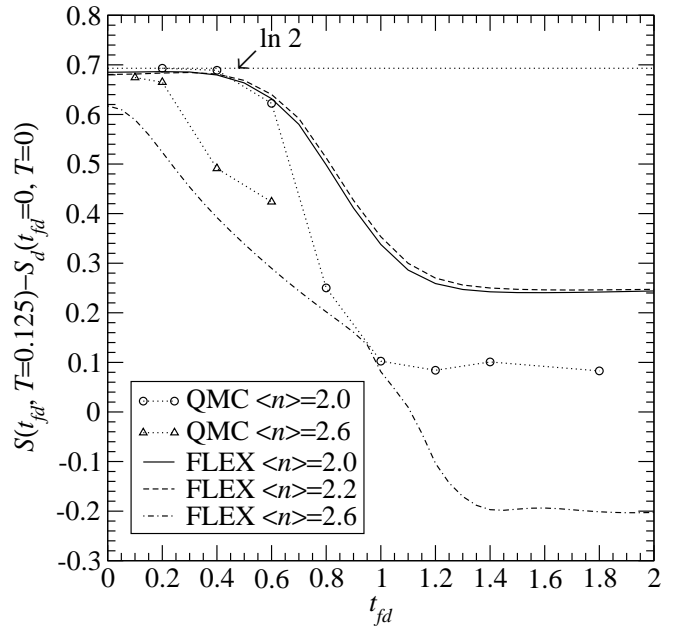}
\vskip 8 pt
\caption{The dependence of the entropy on the
hybridization $t_{fd}$.  Both the QMC and FLEX results are shown.
Parameters are the same as in Fig.~\protect\ref{fig_localmoment}(a),
and the $t_{fd}=0$ entropy of the $d$ system is removed as
discussed in the text.  The QMC and FLEX results agree rather well,
especially in the atomic ($t_{fd}=0$) limit where they produce the $\ln
2$ entropy due to the formation of local moments.}
\label{fig_entropy}
\end{figure}

In Fig.~\ref{fig_entropy} we show the dependence of the low-$T$ entropy
on the hybridization $t_{fd}$ at $T=0.125$.  In order to make the
analysis of the entropy more meaningful, we plot
$S(t_{fd},T)-S_{d}(t_{fd}=0,T=0)$ as a function of $t_{fd}$.
After this subtraction the entropy curves in Fig.~\ref{fig_entropy}
start with the value $\ln 2$ at $t_{fd}=0$, reflecting the
free $f$ moments.  The FLEX results are slightly off, representing the
weak-coupling nature of the approximation, but it is very interesting
to see such a weak-coupling diagrammatic approximation doing so well at
the strong-coupling limit, capturing all of the local-moment
entropy at the lower fillings and 90\% at the highest filling.
For $\langle n \rangle=2.0$ the QMC entropy starts to drop
around $t_{fd}=0.6$; this value moves down to a smaller value of
$t_{fd}=0.2$ for $\langle n \rangle=2.6$.
One note should be added at this point; as the  entropy drop for $\langle 
n \rangle=2.6$ took place at a smaller $t_{fd}$, we didn't pursue 
calculations for higher $t_{fd}$ values.  From Eq. \ref{entropy} 
it is clear that a thin grid of temperatures is needed to get the 
entropy  for a given $t_{fd}$. We were not able to 
reach the needed low temperatures due to ``sign problems'', as mentioned 
in section
\ref{sec_qmc}. 
The drop in the entropy is
also sharper in the $\langle n \rangle=2.0$ case. The $\langle n
\rangle=2.2$ results (only shown for FLEX) are almost identical to the
$\langle n \rangle=2.0$, as seen in other quantities as already noted.
The FLEX and QMC results still don't fully agree as one approaches
$t_{fd}=2$ (weaker coupling) because FLEX becomes truly valid only for
$t_{fd}\gtrsim 3$, as we will show in the next section.

Examination of the singlet correlator (Fig.~\ref{fig_singletcorr}) and
antiferromagnetic structure factor (Fig.~\ref{fig_sf}) suggests that
the sharp drop in the entropy for the $\langle n \rangle=2.0$ and $2.2$
cases could be related to both the formation of the Kondo singlets and the
onset of the antiferromagnetism.  Magnetic order, however, is ruled out
for the $\langle n \rangle=2.6$ case, where there is nevertheless still
a significant drop in the entropy with increasing $t_{fd}$.

\subsection{Internal energy}
\label{sec_internalenergy}

In this section we present results for the internal energy as a
function of $t_{fd}$ at the relatively low temperature of $T=0.125$
(in one case 0.167).  Because the one-body part of the Hamiltonian
dominates the $t_{fd}$ dependence of the internal energy, we report
our results here as differences from the Hartree-Fock (HF)---either
the paramagnetic (PMHF) solution or the broken-symmetry
antiferromagnetic (AFHF) solution---in order to better expose the
more interesting effects of correlations.  This also makes sense
given that the HF solutions must be upper bounds on the correct
internal energy at $T=0$.  Moreover, we might expect the PMHF
solution to be correct in the weak-coupling limit
($t_{fd}\rightarrow\infty$) and have observed previously that the
AFHF energy is correct in the strong-coupling limit ($t_{fd}
\rightarrow 0$) at
low temperature.\cite{ourprl1,lowT} With these two limits in mind,
we first present results as a function of the ratio
$t_{fd}/(t_{fd}+1)$, where values $0$ and $1$ correspond to the
strong- and weak-coupling limits, respectively.

\begin{widetext}
\begin{figure}[hbtp]
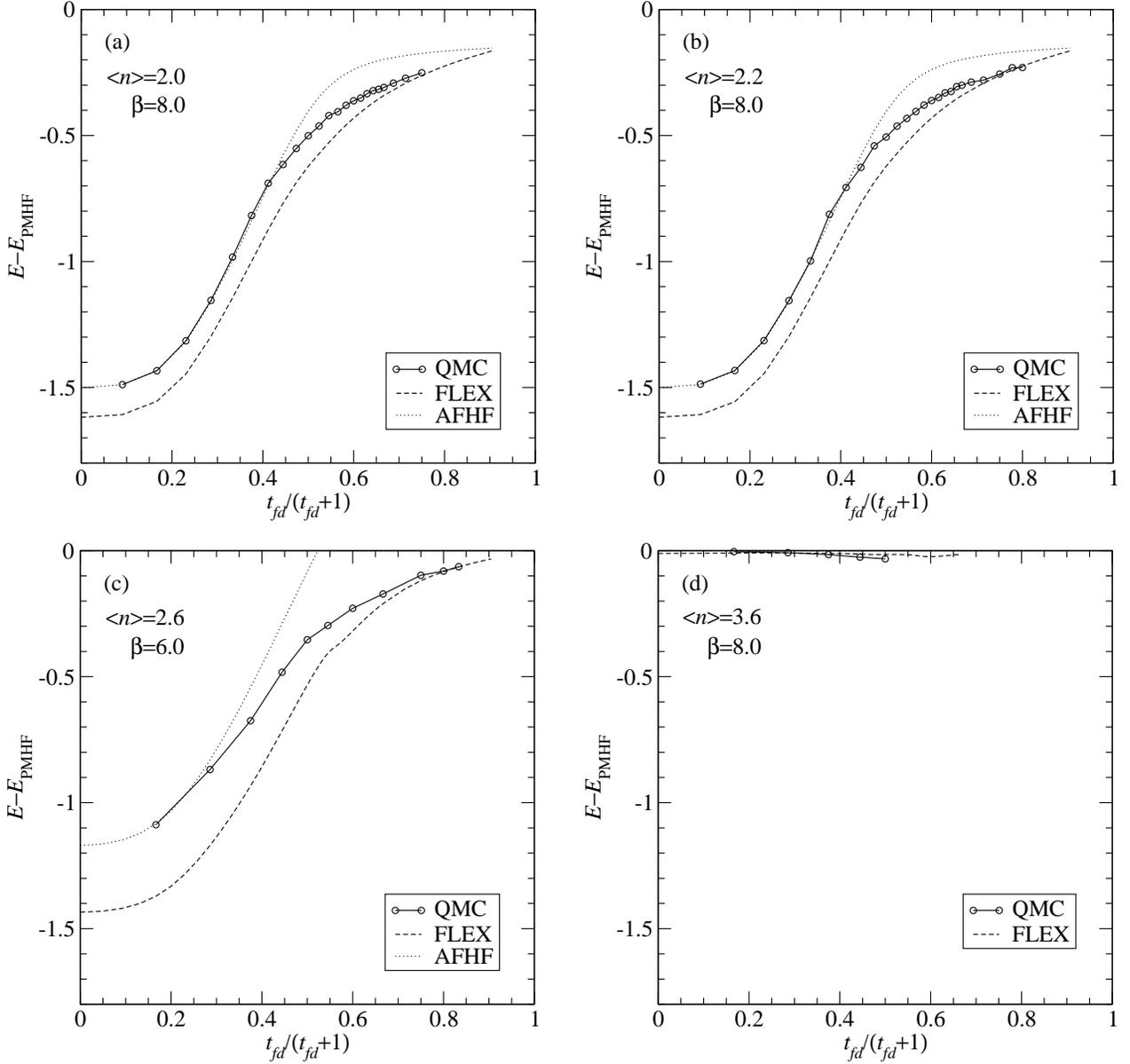

\begin{picture}(0.00,6.90)
\includegraphics{fig6a.eps}
\includegraphics{fig6b.eps}
\includegraphics{fig6c.eps}
\includegraphics{fig6d.eps}
\end{picture}
\caption{The difference of the internal energy for various solutions
from the PMHF solution, $E-E_{\rm PMHF}$, as a function of
$t_{fd}/(t_{fd}+1)$, which is 0 and 1 at the strong- and weak-coupling
limits respectively.  The results for QMC, FLEX, and AFHF are shown.
(a)~$\langle n\rangle=2.0$, (b)~$\langle n\rangle=2.2$, (c)~$\langle
n\rangle=2.6$, (d)~$\langle n\rangle=3.6$. In (d)
$\varepsilon_{f}=-5.4$, and in (c) $\beta=6$.  Other parameters are the
same as in Fig.~\protect\ref{fig_localmoment}(a).  In all cases QMC
agrees with AFHF toward the strong-coupling limit and with FLEX toward
the weak-coupling limit. In the rather extreme filling case of (d), the
energy difference $E-E_{\rm PMHF}$ is two orders of magnitude smaller
than the other cases, implying that the correlations are negligible 
and the PMHF solution is quite accurate.  In general at the
weak-coupling limit, PMHF gives the right solution except when the AFHF
solution has a lower energy than the PMHF as in (a) and (b), which can
happen for small finite systems; then, AFHF gives the right solution at
the weak- as well as the strong-coupling limit.}
\label{fig_epm}
\end{figure}
\end{widetext}

Figure \ref{fig_epm} presents our results for the correlation energy
$E-E_{\rm PMHF}$ as a function of $t_{fd}/(t_{fd}+1)$ at four-different
fillings:  $\langle n \rangle=2.0$ (a), $2.2$ (b), $2.6$ (c), and
$3.6$ (d).  We calculate $E$ with three different methods: QMC,
diagrammatic FLEX, and mean-field AFHF\@.  In the
strong-coupling (atomic) limit, AFHF gives the correct energy at
every filling, agreeing with  QMC.\cite{lowT} Toward the
weak-coupling limit, FLEX
agrees with QMC very well, giving not only the correct
weak-coupling limit itself but also the correct leading-order
dependence on $1/t_{fd}$. The PMHF solution is, in
general, correct {\em at\/} the weak-coupling limit but as is evident
from Fig.~\ref{fig_epm}(c), approaches this limit with an incorrect
dependence on $1/t_{fd}$.

There is an apparent constant offset between
the PMHF and other curves, including AFHF, at the weak-coupling limit
in Figs.~\ref{fig_epm}(a) and (b).  Here one is at half filling or
very close to it, and the AFHF solution on small lattices is lower in
energy at the weak-coupling limit  than the PMHF.
In the thermodynamic limit, these offsets vanish,
which is also suggested in the FLEX size dependence tests
to be cited shortly.

In Fig.~\ref{fig_epm}(d) there is no stable AFHF (or other
magnetic) solution at any coupling strength.  In addition
$E-E_{\rm PMHF}$ is very small compared to (a), (b), and (c).
Indeed correlations are not
important in this case of extreme filling.
Note that while the {\em total\/} fillings
$\langle n \rangle$ vary in Figs.~\ref{fig_epm}(a), (b), and (c), in
each case $\langle n_f \rangle\sim1$ and the correlation energies are
of comparable size.  In sharp contrast $\langle n_f \rangle\sim1.8$ in
Fig.~\ref{fig_epm}(d), confirming the general understanding that
correlations are largest near half filling of an interacting orbital and
diminish for fillings of this orbital away from half filling.

The FLEX approximation is always paramagnetic
if one does not include the anomalous-self-energy diagrams.
Therefore, unlike the mean-field treatments
of correlations, FLEX has no anomalies arising from the solution switching 
from a broken-symmetry (e.g., antiferromagnetic) to the uniform (paramagnetic)
state as occurs in Fig.~\ref{fig_epm}(c).
Such HF transitions are absent in Figs.~\ref{fig_epm}(a) and (b) since the
AFHF solution is always stable.


\pagebreak

\begin{narrowtext}
\end{narrowtext}

\mbox{}

\pagebreak

\mbox{}

\pagebreak

\begin{figure}[hbtp]
\psfig{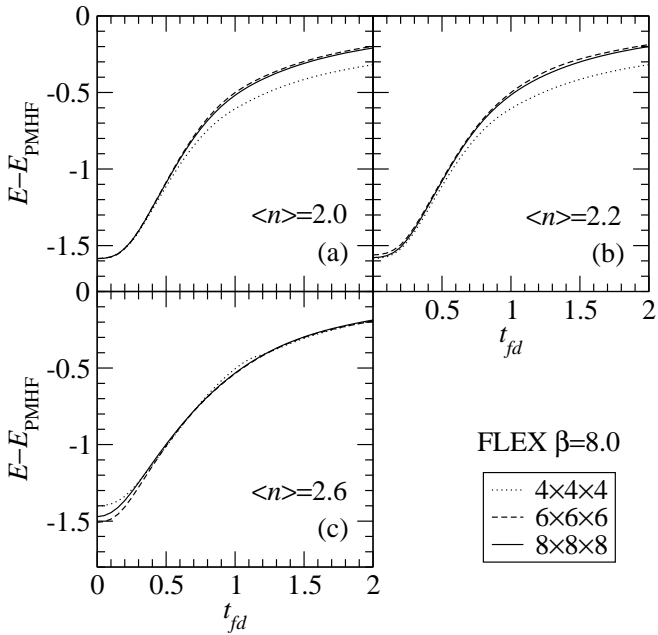}
\vskip 8 pt
\caption{The difference of the FLEX internal energy from the PMHF
internal energy, plotted as a function of the hybridization $t_{fd}$
for different lattice sizes for the purpose of evaluating the
finite-size effects.  (a)~$\langle n\rangle=2.0$, (b)~$\langle
n\rangle=2.2$, (c)~$\langle n\rangle=2.6$.  Parameters are the same as
in Fig.~\protect\ref{fig_localmoment}.  These finite-size effects are
generally small.}
\label{fig_sizeeffects}
\end{figure}

The results in Fig.~\ref{fig_epm} were for $4\times 4\times 4$
periodic cells.  To illustrate the effects of size dependence,
Fig.~\ref{fig_sizeeffects} shows $E_{\rm FLEX}-E_{\rm PMHF}$
calculations done for $4\times 4\times 4$, $6\times 6\times 6$, and
$8\times 8\times 8$ periodic cells. As expected the finite-size
effects on the internal energy are generally small. This is because
the internal energy is basically a local quantity.  Note, however,
the upward shift in $E_{\rm FLEX}-E_{\rm PMHF}$ with increased cell
size for larger $t_{fd}$ in Figs.~\ref{fig_sizeeffects}(a) and (b),
which correspond to Figs.~\ref{fig_epm}(a) and (b), respectively.
These shifts are in the direction to remove the constant offsets at
the weak-coupling limit between the PMHF and more rigorous methods
in Figs.~\ref{fig_epm}(a) and (b).  
It is also the case that the
lowest $T=0$ HF solution at the weak-coupling limit 
is PMHF for all fillings in the thermodynamic limit.

\subsection{Crossover behavior in the PAM}
\label{sec_crossover}

We now discuss some of the implications of these results for the
potential presence of sharp switches in the PAM energy
which could drive phase transitions in appropriate systems.

In the recent work\cite{ourprl1} for the PAM at half filling, $\langle
n\rangle=2.0$, we noted coincident anomalous behavior in three distinct
low-temperature quantities as $t_{fd}$ was increased beyond about
$0.6$:  the singlet correlator increased abruptly; a Kondo resonance
appeared and grew at the Fermi level; and the internal energy broke
away from the AFHF result it had closely followed at smaller $t_{fd}$.
We suggested that the abruptness of this change in internal energy,
evidently associated with the screening of the local moments by the
conduction electrons, was not inconsistent with the physics of the
volume-collapse transitions in the rare-earth metals.  Such a
possibility also was previously discussed in the context of the
Anderson impurity model.\cite{ALLEN}  We now revisit this issue from a
different perspective than given in our earlier work.\cite{ourprl1}

\begin{figure}[hbtp]
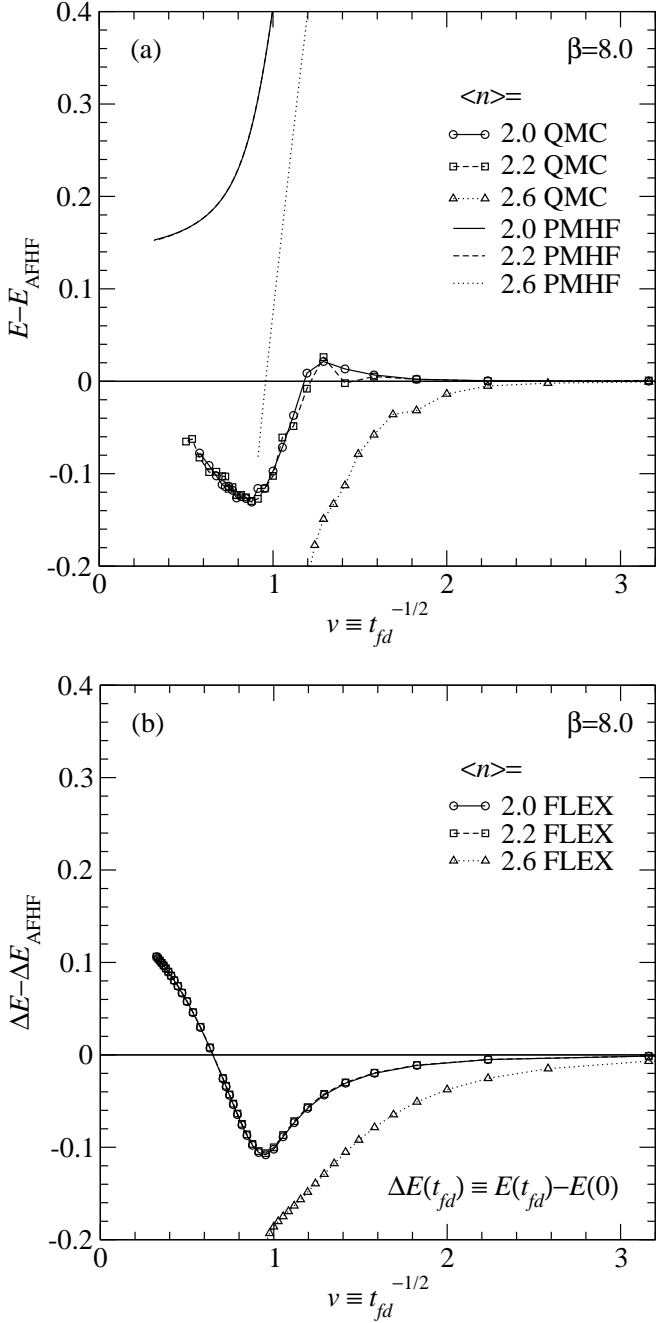

\psfig{figure=fig8a.eps,width=246 pt,angle=0}
\vskip 8 pt
\psfig{figure=fig8b.eps,width=246 pt,angle=0}
\vskip 8 pt
\caption{(a)~The dependence of the QMC and PMHF internal energies
relative to the AFHF as a function of the inverse-square-root hybridization
$v\equiv t_{fd}^{-1/2}$.  Parameters are the same as in
Fig.~\protect\ref{fig_localmoment}(a).  For all three fillings, the
QMC internal energy agrees with the AFHF internal energy at large
$v$, breaking away at an intermediate $v$ value.  This breakaway,
which happens at $v\sim 1.3$ ($t_{fd}\sim0.6$) for $\langle
n\rangle=2.0$ and $2.2$, and at $v\sim 2.2$ ($t_{fd}\sim 0.2$) for
$\langle n\rangle=2.6$, might have implications on the
volume-collapse phenomenon.  (b)~The FLEX results for the same
parameters, with the atomic-limit ($t_{fd}=0$) offsets removed.
The agreement with QMC is quite reasonable although the breakaway
is more gradual than seen for the QMC results in (a).}
\label{fig_eaf} \end{figure}

Figure~\ref{fig_eaf}(a) shows our QMC and PMHF internal energies
relative to the AFHF results, $E-E_{\rm AFHF}$, for all three
fillings at the low temperature of $T=0.125$.  Since the physical
dependence of the hybridization on volume is $t_{fd}\sim V^{-2}$,
we have chosen $v\equiv t_{fd}^{-1/2}\propto V$ to label our
horizontal axis.  Note first the AFHF to PMHF transition at
$v=0.96$ ($t_{fd}=1.09$) which occurs as $v$ is decreased for the
filling $\langle n \rangle=2.6$.  While there are no HF transitions
at the two other fillings for the $4\times 4\times4$ cell, such
transitions do occur in the thermodynamic limit at $v=0.87$ and
$0.74$ ($t_{fd}=1.33$ and $1.83$) for $\langle n \rangle=2.6$ and
$2.2$, respectively, but there is none exactly at $\langle n
\rangle=2.0$.  These mean-field calculations describe a transition
from a ``localized'' AFHF phase, with the $f$ bands Hubbard split
away from the Fermi level, to an ``itinerant'' PMHF phase, with the
merged $f$ bands overlapping the Fermi level, in fundamentally the
same way as realistic HF and modified local-density methods
describe the volume-collapse transitions in the rare-earth
metals.\cite{CEVC}  While having some degree of validity, it has
been argued in Ref.~\onlinecite{CEVC} that the realistic HF and
LDA$+$U transitions occur too close to the itinerant limit for the
rare earths, corresponding to too-small values of $v$ in the
present PAM analog.  The QMC results in Fig.~\ref{fig_eaf}(a)
appear to support this suggestion.\cite{trotter}
While both the AFHF and PMHF
values must be upper bounds on the exact QMC at $T=0$, it is
interesting that at large $v$ the QMC values are quite close to the
AFHF upper bound, whereas at smaller $v$ they appear to bend and
stay well away from the downward-moving PMHF bounds.  The net
effect is a downturn in the QMC energies with decreasing $v$ well
before the PMHF curves cross below the AFHF, as evident for
$\langle n\rangle=2.6$ and likely to be so in the thermodynamic
limit for $\langle n\rangle=2.2$.   This is at least suggestive of
the correlated version of the collapse transition, although, a
compelling case is hindered by the internal-energy contributions
omitted by the PAM.

It should also be emphasized that as $v$ is decreased, the onset of
the downturns in the QMC energies in Fig.~\ref{fig_eaf}(a), $v\sim
1.3$, $1.3$, and $2.2$ ($t_{fd}\sim0.6$, $0.6$, and $0.2$) for
$\langle n\rangle=2.0$, $2.2$, and $2.6$, respectively, coincide
with the onset of rapid increases in magnitude of the singlet
correlator and with similar rapid decreases in the low-temperature
entropy as discussed earlier.  Moreover, from the $\langle
n\rangle=2.6$ case, we know that all these features occur and
coincide in the absence of antiferromagnetic correlations, so that
their fundamental origin is likely associated solely with the
screening of the local moments by the conduction electrons.

As noted earlier the FLEX approximation does not produce the Mott
gap and therefore may overestimate Kondo-like excitations for smaller
$t_{fd}$, which never disappear from the spectrum, even at the
atomic (strong-coupling) limit.  This failure of FLEX leads to an
underestimation of the internal energy as seen at $t_{fd}=0$ in
Fig.~\ref{fig_epm}, due to the spurious presence of Kondo-like
excitations at this limit.  Nevertheless, it is interesting to
focus only on the $t_{fd}$ dependence and remove these atomic-limit
offsets by using $\Delta E(t_{fd})\equiv E(t_{fd})-E(0)$ and
plotting $\Delta E - \Delta E_{\rm AFHF}$ in the case of FLEX, as
is done in Fig.~\ref{fig_eaf}(b).  As may be seen, the agreement
with Fig.~\ref{fig_eaf}(a) is really quite good, although, the
breakaway from the AFHF is more gradual than in the QMC.  It is
tempting to attribute this more gradual behavior in
Fig.~\ref{fig_eaf}(b) to the persistence of Kondo-like excitations
in FLEX to the largest $v\equiv t_{fd}^{-1/2}$, whereas presumably
such excitations cease for the QMC results in Fig.~\ref{fig_eaf}(a)
beyond some critical values of $v$, possibly where the QMC
and AFHF curves become essentially degenerate.

\section{SUMMARY AND CONCLUSIONS}
\label{sec_sum}

In this paper we have extended our previous studies\cite{ourprl1} of
the periodic Anderson model (PAM) to nonsymmetric cases, examining the
low-temperature dependence of a variety of quantities as a function of
the hybridization $t_{fd}$ between the $f$ and $d$ orbitals:  local
moment, singlet correlator, antiferromagnetic structure factor,
entropy, and internal energy.  Results
from the determinant quantum Monte Carlo (QMC), 
fluctuation-exchange approximation (FLEX), and
Hartree Fock (HF) methods have been compared.

Beyond exploration of the properties of the
nonsymmetric PAM itself, these results provide further
insight into the effect of more exact treatments of
correlations on mean field approximations to
phase transitions driven by changes to
the ratio of atomic hybridization to Coulomb repulsion,
such as the volume-collapse transitions in the rare-earth metals.  Indeed
this problem has also motivated many of our parameter choices for the
PAM, which were generally selected to maintain approximately one $f$
electron per site while varying the number of $d$ electrons and
consequently the total filling $\langle n \rangle$.

One of the fundamental results of this work is the coincidence of
systematic changes in the measured quantities across particular values
of the hybridization, $t_{fd}\sim0.6$, $0.6$, and $0.2$ for
$\langle n \rangle = 2.0$, $2.2$, and $2.6$, respectively.  For
$t_{fd}$ less than these values, the local moment ($\sim 1$), singlet
correlator ($\sim 0$), and $f$-subsystem entropy ($\sim \ln
2$) are all approximately constant, while the internal energy is well
approximated by AFHF, i.e., a regime of well-formed, unscreened
local moments.  For $t_{fd}$ greater than these values, the
local moment and entropy drop, the magnitude of the singlet correlator
grows, and the internal energy falls below that of AFHF.  The singlet
correlator suggests Kondo-like effects in this regime, although, there
is a complete absence of antiferromagnetic correlations only at the
largest filling, $\langle n \rangle=2.6$, where this identification
appears unambiguous.  It should be acknowledged that changes with
$t_{fd}$ for $\langle n \rangle=2.6$ are also somewhat more gradual
than those closer to half filling, and for this reason we used slightly
different criteria for the crossover $t_{fd}$ values: the first
significant breakaway from the constant values for $\langle
n\rangle=2.6$ versus a location of particularly rapid change for
$\langle n\rangle=2.0$ and $2.2$.

We have presented the correlation energy $E-E_{\rm PMHF}$ most of the
way between the strong- and weak-coupling limits by varying $t_{fd}$
from 0 to very large values.  The exact QMC energies interpolate
smoothly between the AFHF result at the strong- and PMHF result at
the weak-coupling limit for $\langle n\rangle=2.6$, and we expect
similar behavior at other fillings in the thermodynamic limit.
Approximate diagrammatic FLEX agrees very well with QMC towards weak
coupling and is distinctly improved over PMHF by not only reaching the
correct weak-coupling limit but including the leading order dependence
on $t_{fd}^{-1}$ away from this limit.
Moreover, the FLEX curve is smooth like QMC and generally of the same
shape throughout the full $t_{fd}$ range, although, it increasingly
underestimates the internal energy as the strong-coupling limit is
approached since the FLEX approximation does not produce the Mott
gap.  Even so this is a gradual effect and the $t_{fd}$ dependence
compares favorably to QMC even within the strong- and intermediate-coupling
regimes, $0\leq t_{fd} \leq 2$.  In general the FLEX results
are in remarkably good overall agreement with QMC, suggesting this
approach as a useful approximation for the study of $f$-electron
systems.

Due to the crudeness of the PAM model, it is difficult to make a
compelling argument on thermodynamic grounds that it exhibits some
transition like the volume-collapse transition in the rare-earth
metals.  However, mean-field HF calculations for the PAM and for
realistic all-valence-orbital treatments of the rare earths both
yield transitions of the same fundamental nature, and the latter
are certainly caricatures of the volume-collapse transitions.  It
thus makes sense to ask how the correlated and HF results differ
for the PAM and to presume similar differences should be obtained in
the realistic case.  We find that for decreasing $t_{fd}^{-1/2}$ (or,
volume) the correlated energies break away from the ``localized''
AFHF results at values of $t_{fd}^{-1/2}$ (or, volume) considerably
larger than where the ``itinerant'' PMHF energies cross down
through the AFHF curve and become stable.  Moreover, this breakaway
coincides with the decrease in the local moment and $f$-subsystem
entropy, and the increase in the singlet correlator.  The coupling
of the thermodynamics with the singlet correlator behavior is
certainly consistent with the Kondo volume-collapse
scenario,\cite{ALLEN} while it is interesting that the associated
decay in the local moment is reminiscent of the Mott-transition
perspective.\cite{johansson}

\acknowledgments

We would like to thank Wei Ku for useful discussions and a critical
reading of the manuscript.  Work at University of California, Davis,
was supported in part by Accelerated Strategic Computing Initiative
Grants, IUT Agreement Nos.~B507147 and B511275 with Lawrence Livermore
National Laboratory (LLNL), and by
NSF-DMR-0312261 and NSF-INT-0203837.  T.P. would also like to acknowledge
CNPq-Brazil for financial support.  Work at LLNL was performed under
the auspices of the U.S. Department of Energy by the University of
California, Lawrence Livermore National Laboratory,
under contract No.~W-7405-ENG-48.
We also thank the Materials Research Institute at LLNL for their support.


%
%

%
%


\begin{references}

\bibitem[*]{address_TP}
E-mail address: tclp@if.ufrj.br

\bibitem[\dagger]{address_GE}
E-mail address: esirgen@usc.edu

\bibitem{GENPAM}
G.~R. Stewart, Rev.\ Mod.\ Phys.\ {\bf 56}, 755 (1984); P.~A. Lee,
T.~M. Rice, J.~W. Serene, L.~J. Sham, and J.~W. Wilkins,
Comments Cond.\ Matt.\ Phys.\ {\bf 12}, 99 (1986);
A.~C. Hewson, {\em The Kondo Problem to Heavy Fermions\/}
(Cambridge University Press, Cambridge, 1993).

\bibitem{doniach} S. Doniach, Physica B {\bf 91}, 231 (1977).


\bibitem{RKKY}
M.~A. Ruderman and C. Kittel, Phys.\ Rev.\ {\bf 96}, 99 (1954).

\bibitem{gubernatis2003} 
C. D. Batista, J. Bon\v{c}a, and J. E. Gubernatis,  cond-mat/0308002,
to be published in \prb.

\bibitem{NOZIERES}
P. Nozi\`eres, Ann. Phys. (Paris) {\bf 10}, 19 (1985).

\bibitem{RECENTEXHAUST}
A.~N. Tahvildar-Zadeh, M. Jarrell, T. Pruschke, and J.~K. Freericks,
Phys.\ Rev.\ B {\bf 60}, 10782 (1999).
N.~S. Vidhyadhiraja, A.~N. Tahvildar-Zadeh, M. Jarrell,
and H.~R. Krishnamurthy,
Europhys.\ Lett.\ {\bf 49}, 459 (2000).


\bibitem{Affleck98} V. Barzykin and I. Affleck, \prb {\bf 57},
432 (1998).

\bibitem{Affleck2001} I. Affleck and P. Simon, \prl {\bf 86},
2854 (2001).

\bibitem{gan} J. Gan, J. Phys.: Condens. Matter {\bf6}, 4547 (1994). 

\bibitem{boyce} J. P. Boyce and C. P. Slichter, \prl {\bf 32}, 61
(1974); \prb {\bf 13}, 379 (1996).


\bibitem{ourprl2}
K. Held, C. Huscroft, R.~T. Scalettar, and A.~K. McMahan,
Phys.\ Rev.\ Lett.\ {\bf 85}, 373 (2000).


\bibitem{qmcpam}
Two-dimensional quantum Monte Carlo (QMC) studies of the PAM include:
Y. Zhang and J. Callaway,
Phys.\ Rev.\ B {\bf 38}, 641 (1988);
M. Vekic, J.~W. Cannon, D.~J. Scalapino, R.~T. Scalettar, and R.~L. Sugar,
Phys.\ Rev.\ Lett.\ {\bf 74}, 2367 (1995).
In three dimensions see Ref.~\protect\onlinecite{ourprl1}.

\bibitem{cpqmcpam}
A two-dimensional QMC study of
the PAM using the ``constrained-path'' method
to reach low temperatures is contained in
J. Bon\u{c}a and J.~E. Gubernatis,
Phys.\ Rev.\ B {\bf 58}, 6992 (1998).

\bibitem{DMFTPAM}
Dynamical mean-field theory studies of the PAM include:
M. Jarrell, Phys.\ Rev.\ B {\bf 51},
7429 (1995); A.~N. Tahvildar-Zadeh, M. Jarrell, and J.~K. Freericks,
{\em ibid.}\ {\bf 55}, 3332 (1997); M. Rozenberg,
{\em ibid.}\ {\bf 52}, 7369 (1995);
T. Schork and S. Blawid,
{\em ibid.}\ {\bf 56}, 6559 (1997);
T. Pruschke, R. Bulla, and M. Jarrell,
{\em ibid.}\ {\bf 61}, 12799 (2000);
D. Meyer and W. Nolting,
{\em ibid.}\ {\bf 62}, 5657 (2000);
{\bf 61}, 13465 (2000);
Y. Shimizu, O. Sakai, and A.~C. Hewson,
J. Phys.\ Soc.\ Jpn.\ {\bf 69}, 1777 (2000);
K. Held and R. Bulla,
Eur.\ Phys.\ J. B {\bf 17}, 7 (2000).

\bibitem{WHY}
The physical motivation for the choice of
on-site hybridization is open to question since
one can always choose an orbital basis in which
the one-particle Hamiltonian is diagonal at a given
lattice site.

\bibitem{ALLEN}
J.~W. Allen and R.~M. Martin, Phys.\ Rev.\ Lett. {\bf 49} 1106, (1982);
J.~W. Allen and L.~Z. Liu, Phys.\ Rev.\ B {\bf 46}, 5047 (1992);
M. Lavagna, C. Lacroix, and M. Cyrot,
Phys.\ Lett.\ A {\bf 90}, 210 (1982); J. Phys.\ F {\bf 13}, 1007 (1983).

\bibitem{johansson}
B. Johansson, Philos.\ Mag.\ {\bf 30}, 469 (1974);
B. Johansson and A. Rosengren, Phys.\ Rev.\ B {\bf 11}, 2836 (1975);
B. Johansson, I.~A. Abrikosov, M. Ald\'en, A.~V. Ruban, and H.~L. Skriver,
Phys.\ Rev.\ Lett.\ {\bf 74}, 2335 (1995).

\bibitem{CEVC}
A. McMahan, C. Huscroft, R.~T. Scalettar, and E.~L. Pollock,
J. of Computer-Aided Materials Design {\bf 5}, 131 (1998).

\bibitem{andyprb}
A. K. McMahan, K. Held, and R. T. Scalettar,
Phys.~Rev.~{\bf B67}, 075108 (2003).


\bibitem{ourprl1}
C. Huscroft, A.~K. McMahan, and R.~T. Scalettar,
Phys.\ Rev.\ Lett.\ {\bf 82}, 2342 (1999).

\bibitem{HIRSCH}
J.~E. Hirsch and S. Tang, Phys.\ Rev.\ Lett.\ {\bf 62}, 591 (1989).

\bibitem{SRW}
S.~R. White, D.~J. Scalapino, R.~L. Sugar,
E.~Y. Loh, J.~E. Gubernatis, and R.~T. Scalettar,
Phys.\ Rev.\ B {\bf 40}, 506 (1989).

\bibitem{COMMENT}
This decreased sensitivity of long-range AF order to doping is
almost certainly tied to the fact that in the two--band model
we keep $\langle n_f \rangle=1$ and are doping the conduction band.

\bibitem{motthubbardgap}
Our extensive studies in recent years
indicate that the FLEX approximation does not produce
the Mott-Hubbard gap in general (except for certain
parameters). But there is usually a hump structure
reminiscent of the Hubbard bands, which are caused
by the strong spin fluctuations, surrounding the central
Kondo-like resonance. The more complicated diagrammatic parquet
approximation also fails to generate the Mott-Hubbard gap
[N.~E. Bickers (unpublished)].

\bibitem{LDADMFT}
For treatments which incorporate first-principles band structures
and interaction strengths, see
M.~B.~Z\"olfl,
I.~A.~Nekrasov, Th.~Pruschke, V.~I.~Anisimov, and J.~Keller,
Phys.\ Rev.\ Lett.\ {\bf 87}, 276403 (2001); and
K. Held, A. McMahan, and R.~T. Scalettar,
 Phys.\ Rev.\ Lett.\ {\bf 87}, 276404 (2001).

\bibitem{DETQMC}
R. Blankenbecler, R.~L. Sugar, and D.~J. Scalapino,
Phys.\ Rev.\ D {\bf 24}, 2278 (1981).

\bibitem{LOH}
E.~Y. Loh, J.~E. Gubernatis, R.~T. Scalettar, S.~R. White,
D.~J. Scalapino, and R.~L. Sugar, Phys.\ Rev.\ B {\bf 41}, 9301 (1990).

\bibitem{FLEX}
N.~E. Bickers and D.~J. Scalapino, Ann.\ Phys.\ (N.Y.) {\bf 193}, 206 (1989).

\bibitem{baymkadanoff}
G. Baym and L.~P. Kadanoff, Phys.\ Rev.\ {\bf 124}, 287 (1961);
G. Baym, {\em ibid.}\ {\bf 127}, 1391 (1962).

\bibitem{PARQUET}
C. de~Dominicis and P.~C. Martin, J. Math.\ Phys.\ {\bf 5}, 14 (1964).

\bibitem{SCFLEX}
C.-H. Pao and N.~E. Bickers, Phys.\ Rev.\ Lett. {\bf 72}, 1870 (1994);
P. Monthoux and D.~J. Scalapino, {\em ibid.}\ {\bf 72}, 1874 (1994).

\bibitem{MOREFLEX}
N.~E. Bickers, D.~J. Scalapino, and S.~R. White, Phys.\ Rev.\ Lett.\ {\bf 62},
961 (1989); N.~E. Bickers and S.~R. White, Phys.\ Rev.\ B {\bf 43}, 8044
(1991); G. Esirgen and N.~E. Bickers, {\em ibid.}\ {\bf 55}, 2122 (1997);
{\bf 57}, 5376 (1998); G. Esirgen, H.-B. Sch\"{u}ttler, and
N.~E. Bickers, Phys.\ Rev.\ Lett.\ {\bf 82}, 1217 (1999);
G. Esirgen, H.-B. Sch\"{u}ttler, C. Gr\"{o}ber, and H.~G. Evertz,
 Phys.\ Rev.\ B.\ {\bf 64}, 195105 (2001).

\bibitem{paorg}
C.-H. Pao and N.~E. Bickers, Phys.\ Rev.\ B {\bf 49}, 1586 (1994).

\bibitem{vertex}
The magnetic susceptibilities plotted are the
FLEX susceptibilities (bubble-diagram sums) which go directly into the
FLEX self-energy. They are {\em not\/} the conserving FLEX
susceptibilities calculated by the functional differentiation of the
generating functional. The extra diagrams which enter the conserving
susceptibilities tend to increase the magnitude of the magnetic
susceptibilities, but qualitative behavior of the magnetic
susceptibilities remain the same.

\bibitem{2dHub}
T.~Paiva, R.T.~Scalettar, C.~Huscroft, and A.K.~McMahan, Phys. Rev. B
{\bf 63}, 125116 (2001).

\bibitem{lowT}
It is to be emphasized that the AFHF and QMC internal energies
agree in the strong-coupling (small-$t_{fd}$ limit) only at low
temperatures.  The temperature dependence is qualitatively similar
to Fig.~\ref{fig_epm}(c) with the $x$ axis being the temperature,
where the AFHF energy switches to the PMHF at higher temperature
and the QMC interpolates smoothly between the low- and
high-temperature limits.

\bibitem{trotter}
The data shown in Fig.~\ref{fig_eaf}(a) 
differ slightly from Ref.~\protect\onlinecite{ourprl1}.
in that the energy difference has a small upturn before
becoming negative with decreasing volume.  This is
because the present data now includes a correction
for the errors associated with the finite discretization
of inverse temperature.  

\end{references}
\end{document}